# Location Prediction of Social Images via Generative Model

Xiaoming Zhang[1], Zhoujun Li[1], Senzhang Wang[1], Yang Yang[1], Xueqiang Lv[2]

[1]School of Computer Science and Engineering, Beihang University, China
[2]Beijing Key Laboratory of Internet Culture and Digital Dissemination Research, China
{yolixs, lizj, szwang, yangyangfuture }@buaa.edu.cn ; lxq@bistu.edu.cn

## ABSTRACT
The vast amount of geo-tagged social images has attracted great attention in research of predicting location using the plentiful content of images, such as visual content and textual description. Most of the existing researches use the text-based or vision-based method to predict location. There still exists a problem: how to effectively exploit the correlation between different types of content as well as their geographical distributions for location prediction. In this paper, we propose to predict image location by learning the latent relation between geographical location and multiple types of image content. In particularly, we propose a geographical topic model GTMI (geographical topic model of social image) to integrate multiple types of image content as well as the geographical distributions, In GTMI, image topic is modeled on both text vocabulary and visual feature. Each region has its own distribution over topics and hence has its own language model and vision pattern. The location of a new image is estimated based on the joint probability of image content and similarity measure on topic distribution between images. Experiment results demonstrate the performance of location prediction based on GTMI.

## Categories and Subject Descriptors
H.3.1 [**Information storage and retrieval**]: Content Analysis and Indexing. H.2.8 [**Database management**]: Data mining.

## 1. INTRODUCTION
With the rapid development of Web 2.0 and GPS-equipped mobile terminals, geo-tagged social media data are tremendously increasing. These location-based social network (LBSN) services, such as Flick and Google Latitude, not only allow users to maintain cyber links with other users, but also enable users to share their activities happening at certain locations in various forms. Usually, users associate their documents in social network with a geographic record which is mainly denoted by a two dimensional vector, i.e., latitude and longitude. Nowadays, Flickr hosts more than 100 million images associated with textual descriptions (e.g., titles, comments, and tags), visual contents, and GPS records. This large amount of geo-tagged social images presents various popularities across different geographical regions. That is, the language models and vision patterns of social images relate to their locations. For example, images about New York City might cover entirely different events compared to those about Beijing. The preference of tag words and visual contents are different for the two cities. On the other side, there are also many social images remained untagged. These characteristics make it possible to learn a reasonable model to identify the relation between location and distribution of image content including textual description and visual content, which is important for predicting the location of a untagged image.

Recently, many approaches have been proposed to estimate the geographic location of social image [1, 2, 4, 6, 22]. These approaches can be categorized into two classes. The first class of approaches is vision-based, which estimates the location of new image based on the locations of visually similar images [1, 11, 22]. However, due to the large variety of visual content and semantic gap problem [23, 24], exploiting visual content only is challenging. Another class of approaches is text-based, which exploits the relation between text description and location to predict where the image was taken [3, 16]. These approaches perform better than vision-based approaches, since the text words are more effective in conveying the location information. Besides, some location-specific tags (e.g., Summer Palace and Forbidden City) are helpful to disambiguate some visually similar images. However, these approaches mainly employ a pure language model, and thus they are ineffective to process social images with rich visual contents. Unlike textual document, geo-tagged social image contains multiple types of contents, i.e., textual description, visual contents, and geographical information. Each location has its own characteristic of both language model and vision pattern, and different types of content are also correlated with each other. These contents should be incorporated simultaneously in a model to identify relations between social image and location. Although there are some works on using different content for landmark prediction and representation, these works unite different types of feature linearly. There still exists an challenge: how to learn the latent correlation between visual content and text content for each location and in turn estimate the location based on the latent correlation.

To address this problem, we propose a geographical topic model of social image (GTMI), by simultaneously incorporating geographical information, textual description and visual contents. To combine the different data modalities of text content and visual content, we propose a common structure shared by both domains, i.e., image topic. Each topic is modeled as a distribution on both text words and visual features. GTMI could identify different topic patterns across regions, where the geographical characteristics of languages and visual contents are integrated by topics consistently. That is, each region has its own topic distribution, and hence has its own language models and vision patterns. A generative procedure is employed to model the production of text content and visual content, i.e., image patches that comprise of the pixels that are spatially coherent, based on location information. The location of a new image can be predicted by a two-step method. First, the region which has the greatest joint probabilities of the query image's content is selected. Then, the location is estimated by propagating the locations from the most similar images in this region to the query image, and the influences of these similar images are determined by their weights. Experimental results show that our GTMI outperforms non-trivial baselines on predicting image location. Compared with existing works, our main contributions are as follows:

1. We propose a geographically generative model of image content and locations, which incorporates multiple facets of image environments in an integral fashion. A Gibbs sampling method is employed to infer the model parameters.
2. A two-stage strategy is proposed to predict image location based on GTMI, which exploits the latent topic distribution of region and image.
3. A set of experiments are conducted on a real-world social image dataset, and the experimental result shows that the proposed model outperforms several state-of-the-art models on image location prediction.

The remainder of this paper is organized as follows. In the next section, we introduce related works. We describe our model in Section 3, and location prediction based on GTMI is introduced in Section 4. The experiments are described in section 5. Finally, the paper is concluded in section 6.

## 2. RELATED WORKS

The study of predicting the location at which an image was taken has drawn much research attention from the computer vision and data mining community. Most of the studies addressing this task fall into two categories: text-based prediction, which predict the location based on the text content associated with social image using language model, and vision based prediction, which propagates the locations from visually similar images to the query image.

User-contributed text tags have been used as a basis of a large range of successful geo-coordinate predication algorithms. These works exploit the geographical characteristic of language model to mine what tags are location-specific. LGTA (Latent Geographical Topic Analysis) [3] combines geographical clustering and topic modeling to identify the geographical topics of social images, as well as estimate the topic distributions in different geographical locations for topic comparison. Another work proposes a language model based on user annotations, to place the annotated Flickr images on the map [6]. The MDP-based geographical topic model (MGTM) captures dependencies between geographical regions to support the detection of text topics with complex, non-Gaussian distributed spatial structures [15]. This model is based on a multi-Dirichlet process (MDP). In [16], a two-step approach is proposed to estimate where a given image or video was taken, using only the tags that a user has assigned to it. In the first step, a language modeling approach is adopted to find the area which most likely contains the geographic location of the image. Then, a precise location is determined within the area that was found to be most plausible. Those works mainly use a pure language model to identify the link between text content and location, which neglect the relation between text content and visual features as well as their distributions over locations.

Another category of approaches are vision-based. These works predict the location of image based on the visual features of image only. The method IM2GPS first retrieves visually similar images and form clusters based on geo-coordinate information [1]. The geo-centroid of the cluster containing the most images is used as the predicted location. GVR [11] searches a set of candidate images that are visual neighbors of the query image and expands each candidate image with a geo-visual expansion set of images that are geographical neighbors of the candidate. The candidate images are ranked according to the visual similarity of their geo-visual expansion sets and the query image. Then, the locations of the top ranked candidates are propagated to the query image. There are also some vision-based approaches for landmark recognition [17] and scenes matching [18]. The performance of most of these approaches falls short of the performance of text-based approaches in the large-scale location prediction.

Besides, there are also some works using both text description and visual content to mine the link between landmark and image. For example, the content analysis (based on text tags and visual features) is combined with structural analysis (based on geospatial data) for landmark recognition [2]. Similarly, multiple types of contents, i.e., locations, tags and visual features, are used to generate diverse and representative images for landmarks in [5]. However, these approaches consider different types of feature independently, which are not effective in learning the latent relation between different types of contents. In this paper, a novel generative model integrating geographical information, textual description and visual contents with their correlation is proposed to mine the distributions of geographical language and vision patterns across different regions, by which new image locations are predicted.

## 3. GEOGRAPHICAL TOPIC MODEL of SOCIAL IMAGE

Geo-tagged image contains multiple types of content, i.e., visual content, textual description, and geographical information. Usually, the text content and image visual content are highly correlated [19] and they also relate to the location. We propose to identify the language models and vision patterns for each location, based on which location prediction can be developed. However, incorporating those contents simultaneously in the model to identify relations between image content and location is challenging, since the text space and visual content space have inherently different structures. To address this problem, it is necessary to apply a common structure to link them. On the other hand, this structure is also can be used to discover the geographical characteristic of each location. Thus, we propose a geographical image topic model GTMI, in which latent topic is used as the common structure and modeled on both text feature and visual feature. The geographical language models and visual content patterns are reflected by the topic distribution corresponding to the location. We accomplish this thanks to the large amount of social image data and the diversity of language and visual content variations appearing in social images. There are many factors that influence the language and visual content used in a social image taken in a particular location. For example, textual words used in an image depend on the local culture and the visual content of the image, while the visual content depend on the local view of the geographical region, e.g., famous building, nature view, and local sports activity. We will take these factors into the construction of GTMI.

The graphic representation of GTMI is shown in figure.1. In GTMI, geo-coordinates are grouped into regions, and each image belongs to a region which has its own distribution over topics. Each topic is represented by two topic-specific distributions: topic-specific word distribution and topic-specific distribution over visual features. The two topic-specific distributions are correlated with each other. The topic-specific word distribution is modeled as a multinomial distribution, and the topic-specific distribution over visual features is modeled as a normal distribution. Then, the location of a query image is predicted based on the geographical model, i.e., estimating the joint probability of its contents given a region and propagating the location from the most similar images of this region to the query image. The geographical topic model can also be used to other image application, such as image tagging, image retrieval, and image clustering and so on.

## 3.1 The generative procedure

Compared with the text words, visual features, e.g., color, edge, and texture, are much lower level representation on semantics. We introduce a middle-level feature for image representation. Each image is segmented into multiple patches that comprise of the pixels that are spatially coherent and perceptually similar with respect to certain appearances [20]. Then, each patch is represented by a $|f|$-dimensional visual features, e.g., Bag-of-Words (BOW) features. Therefore, each geo-tagged social image $p=\{w_p, f_p, l_p\}_{p\in P}$ consists of three atoms: $w_p$ is a vector of words extracted from its textual contents, e.g., tags, titles, comments, $f_p =(f_{p,1}, f_{p,2}, …, f_{p,N_{f_p}})$ is a set of patches segmented from $p$, and $l_p$ is a real-valued pair $l_p=\{la, lo\}$, representing the latitude and longitude where the image is taken. For simplicity, we assume that all the textual contents in our data are generated by a fixed vocabulary of $W$ words, and the geographical locations are clustered into $R$ latent regions and the topic number is $K$. Each topic $z \in Z$ is generated from regions instead of documents. The notations are list in Table 1.

**Table 1. Notations used in the paper**

| Notation | Size | Description |
|---|---|---|
| $\mu^l$ | $R^2$ | Mean location of a latent region |
| $\Sigma^l$ | $R^{2\times 2}$ | Covariance matrix of a latent region |
| $\mu^f$ | $R^{|f|}$ | Mean visual feature of a latent topic |
| $\Sigma^f$ | $R^{|f|\times|f|}$ | Covariance matrix of a latent topic |
| $\psi$ | $R\times W$ | Region-specific word distribution |
| $\beta$ | $K\times W$ | Topic-specific word distribution |
| $\xi$ | $R\times K$ | Region-specific topic distribution |
| $\tau$ | $R\times 2$ | Region-specific topic type distribution |

The geographical distribution of each region is assumed to be normal $N(\mu^l_r, \Sigma^l_r)_{r:1..R}$, where $\mu^l_r$ and $\Sigma^l_r$ are the mean vector and covariance matrix of region $r$, respectively. Moreover, the topic-specific visual features are also assumed to follow a normal distribution, parameterized as $(\mu^f, \Sigma^f) = \{(\mu^f_k, \Sigma^f_k)\}_{k:1..K}$. The words that are close in space are more likely to belong to the same region, and they are more likely to be generated by the same topics. Similarly, the visual features that are close in space are more likely to appear in the same region, and they are more likely to be clustered into the some topic. Our model GTMI has the following intuitions:

1. Words used in a social image depend on both the location and topic of the image, while the topic generating the words depends on the topic distribution of the location and has influence on the topics of visual contents of this image.
2. Visual features used in a social image depend on the semantic of this image, and therefore the topic of visual feature depend on the topics assigned to the text words corresponding to the image.
3. Topics have different distributions over different regions. Different geographical regions have different language variations and different distributions of visual patterns.

Figure. 1 depicts a graphical representation of GTMI. To generate a geo-tagged image $p$, the generative procedure of GTMI can be described as following:

1. Sampling a region $r$ from the discrete distribution of region importance $\varepsilon$, $r \sim Discrete(\varepsilon)$.

2. Sample location $l_p$ from normal distribution of $N(\mu^l_r, \Sigma^l_r)$.

3. To generate the visual feature of each patch $f_{p,i}$ in image $p$

   i: Sample topic: $z^p_{f_{p,i}} \sim Multinomial(\xi)$

   ii: Sample visual features: $f_{p,i} | z^p_{f_{p,i}} = k \sim N(\mu^f_k, \Sigma^f_k)$.

4. To generate every word $w_{p,i}$ in image $p$

   Sample a coin: $x_i \sim Bernoulli(\tau_r)$.

   If $x_i = 0$

   Sample word: $w_{p,i} \sim Multinomial(\psi_r)$.

   If $x_i = 1$

   i: Sample topic: $z^p_i \sim Multinomial(\xi)$ conditioned on image patches.

   ii: Sample word: $w_{p,i} | z^p_i = k \sim Multinomial(\beta_k)$.

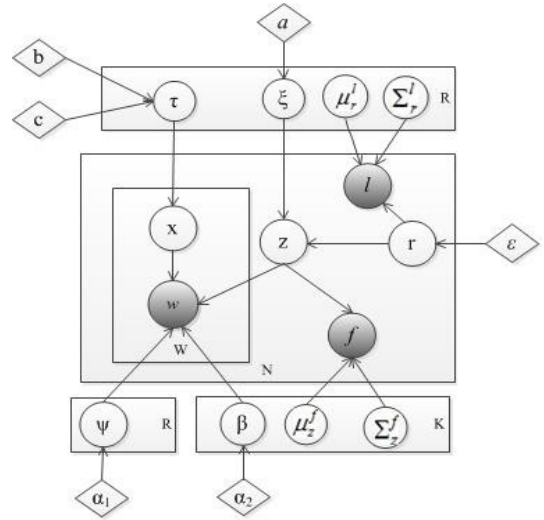

**Figure. 1. A graphical representation of GTMI**

### 3.1.1 Generation of Textual Words

Textual word describes the context or the semantic information of image's visual contents, which suggests that each word can be generated by correlating them to the collective set of topic indicators selected from the image generating region. As different regions have their own language characteristics, we adopt an additional machinery to handle special words, which are similar to the subtraction of document-specific words [9]. Beside the standard latent topic produced by standard topic model [12], we introduce a region-specific topic $\psi$ (sampled from $Dir(\alpha_l)$ once for each region) to generate the region-specific words. Thus, the topics in our model consist of two types, i.e., region topic and standard latent topic which is similar to those produced by the LDA topic model [12].

The generative process for textual words now proceeds as follows. For each image $p$, we associate a Bernoulli distribution $\tau$ with prior parameters $b$ and $c$, which models the distributions of region-specific words, and latent topic words. As shown in step 4 of the generative procedure, to generate a word $w_{p,i}$ of image $p$, we first sample a random variable $x_i$ from a region-specific Bernoulli distribution $\tau_r$, which in turn has prior parameters $b$ and $c$. If $x=0$, the word is sampled from the region-specific topic; if $x=1$, a standard latent topic indicator, $z_i^p$, is selected according to topic distributions on the region and visual feature generation. The intuition is that, if a topic has a high co-occurrence with the region from which the image is sampled and has a high productive probability for the corresponding words and image patches, it has a high probability to be chose.

### 3.1.2 Generation of Visual Features

The visual feature of each image patch is modeled as a normal distribution whose mean and variance are topic-specific. Many works embellish the parameters of a normal distribution with an inverse Wishart prior [13], which are computationally expensive. In this paper, we take a simpler approach by placing a non-informative Jeffrey's prior over the values of the mean parameters, i.e. $\mu_z^f \sim Unif$. Meanwhile, an inverse prior over the variance is placed to penalized large variances, i.e. $P(\Sigma_z^f) \propto (\Sigma_z^f)^{-1}$ [13]. It is because that the calculation of image features might introduce noises. With such prior, the estimation of $\Sigma_z^f$ for a give topic is more robust to outliers. Then, the pdf function for an image, given a topic-specific normal distribution, is revised as a function $t(f;\mu,\Sigma,n)$, which is a student $t$-distribution with mean $\mu$, variance $\Sigma$, and $n$ degree of freedom. Similarly, the pdf function $P(l_p | \mu_r^l, \Sigma_r^l)$ for a geographical location, given a region-specific normal distribution, is also revised as a student $t$-distribution function.

### 3.2 Inference

Under the generative process, we seek to compute the posterior probability:

$$P(\gamma, \psi_{1:R}, \beta_{1:K}, \xi_{1:R}, \mu_{1:R}^l, \Sigma_{1:R}^l, \mu_{1:K}^f, \Sigma_{1:K}^f | a,b,c,\alpha_{0:2}, \mathbf{w}, \mathbf{f}, \mathbf{l}) \quad (1)$$

The above posterior probability can be easily written down from the generative model. However, the posterior is intractable. We approximate it via a collapsed Gibbs Sampling procedure [7, 8], by integrating the hidden variables, e.g., the topic-mixing vectors of each region, the coin base for each region, and the topic distributions over all modalities. Therefore, the state of the sampler at each iteration contains the topic indicators for all regions. We alternate sampling each of these variables conditioned on its Markov blanket until convergence. When it converges, the expected values of all the parameters that were integrated out can be calculated. To simplify the calculation of the Gibbs sampling update equations, we keep a set of sum matrices with the form $C_{xy}^{XY}$ to denote the number of times instance $x$ appeared with instance $y$. Moreover, the subscript $-i$ is used to denote the same quantity it is added to without the contribution of item $i$. For example, $C_{wk}^{WZ}$ denotes the number of times word $w$ as sampled from latent topic $k$, and $C_{wk,-i}^{WZ}$ is the same as $C_{wk}^{WZ}$ without the contribution of word $w_i$. The sampling procedure can be described as following:

For each image $p$, a latent region $r$ is firstly drawn from the following distribution, conditioned on the old topic assignments:

$$r | p, \Phi \sim P(r_j | \varepsilon) P(l_p | \mu_{r_j}^l, \Sigma_{r_j}^l) P(w_p | r_j) P(f_p | r_j) \quad (2)$$

where $P(l_p | \mu_{r_j}^l, \Sigma_{r_j}^l)$ is the pdf function for a multivariate normal distribution corresponding to region $r_j$. $P(r_j | \varepsilon)$, $\mu_r$ and $\Sigma_r$ are estimated as following:

$$\mu_{r_j}^l = \frac{1}{Num(p,r_j)} \sum_{p \in \mathbf{P}} g(r(p) = r_j) l_p \quad (3)$$

$$\Sigma_{r_j}^l = \frac{\sum_{p \in \mathbf{P}} g(r(p) = r_j)(l_p - \mu_{r_j}^l)^T (l_p - \mu_{r_j}^l)}{Num(p,r_j) - 1} \quad (4)$$

$$P(r_j | \varepsilon) = \frac{\sum_{p \in \mathbf{P}} g(r(p) = r_j) + \varepsilon}{|\mathbf{P}| + \varepsilon |\mathbf{R}|} \quad (5)$$

where $Num(p,r_j)$ is the number of images assigned to region $r_j$, and $g(r(p) = r_j)$ is a indicator function which is 1 if and only if the image $p$ is assigned to region $r_j$. The component $P(w_p | r_j)$ is estimated as following:

$$P(w_p | r_j) = \prod_{w_{p,i}} P(w_{p,i} | r_j) = \prod_{w_{p,i}} (P(x_i = 0 | r_j) P(w_{p,i} | \psi_j)$$
$$+ P(x_i = 1 | r_j) \sum_{z \in Z} P(w_{p,i} | z) P(z | r_j)) \quad (6)$$

$$P(x = s | r_j) = \frac{C_{rs}^{RX} + \lambda_s(b,c)}{\sum_{x'} C_{rx'}^{RX} + b + c} \quad (7)$$

$$P(w_{p,i} | \psi_j) = \frac{C_{ir}^{WR} + \alpha_1}{\sum_{w'} C_{w'r}^{WR} + W\alpha_1} \quad (8)$$

$$P(w_{p,i} | z = k) = \frac{C_{ik}^{WZ} + \alpha_2}{\sum_{w'} C_{w'k}^{WZ} + W\alpha_2} \quad (9)$$

$$P(z = k | r_j) = \frac{C_{rk}^{RZ} + a}{\sum_{k'} (C_{rk'}^{RZ} + a)} \quad (10)$$

where $C_{rk}^{RZ}$ denotes the times that topic $k$ is assigned to the word tokens in region $r$, $\lambda_s(b,c) = b$ and $c$ for $s = 0$ and 1 respectively, and the variable $x_i$ act as a switch: if $x_i$=1, the word is generated by the standard topic production mechanism, whereas if $x_i$=0 the word is sampled from a region-specific multinomial. $C_{rs}^{RX}$ counts the number of words in region $r$ is assigned the region-specific topic ($s$=0), and the number of words in region $r$ is assigned a standard latent topic ($s$=2). $P(f_p|r_j)$ is estimated as following:

$$P(f_p|r_j) = \prod_{f_{p,i}} \sum_{k=1}^{K} P(f_{p,i}|z=k)P(z=k|r_j) \quad (11)$$

where $P(f_{p,i}|z=k)=P(f_{p,i}|\mu_k^f,\Sigma_k^f)$ is the productive probability of the visual features, the pdf of a multivariate normal distribution, $\mu_k^f$ is the sample mean of the values of image feature that are assigned to topic $k$, and $\Sigma_k^f$ is defined similarly.

Then, we update the topic assignments for the textual words and visual features of image $p$ conditioned on the region assignment as following:

Sample a topic for every patch $f_{p,i}$:

$$P(z=k|f_{p,i},w_p) \propto \frac{C_{rk}^{RZ}+a}{\sum_{k'}(C_{rk'}^{RZ}+a)} \cdot P(f_{p,i}|z_{f_{p,i}}=k) \quad (12)$$

where the first part measures the comparability of joining a topic, given the region. As described above, the pdf function for a patch, given a topic-specific normal distribution, can be revised as a function $t(f;\mu,\Sigma,n)$ of a student $t$-distribution. Thus, the second part is calculated by:

$$P(f_{p,i}|z=k) \propto t(f;\hat{\mu}_k^f,\hat{\Sigma}_k^f,C_{fk,-i}^{FZ}-1) \quad (13)$$

where $\hat{\mu}_k^f$ is the sample mean of the values of image features assigned to topic $k$, $\hat{\Sigma}_k^f$ is defined similarly, and $C_{fk}^{FZ}$ is the number of times that image patches are sampled from topic $k$. Additionally, the pdf function $P(l_p|\mu_{r_j}^l,\Sigma_{r_j}^l)$ in Eq. (2) can also be calculated similarly.

Sample a topic for every word token $w_{p,i}$ in $p$:

$$P(x_i=0|\mathbf{w},\mathbf{x}_{-i},\mathbf{z}_{-i},\alpha_1,b,c)$$
$$\propto \frac{C_{r0,-i}^{RX}+b}{\sum_{x'} C_{rx',-i}^{RX}+b+c} \cdot \frac{C_{wr,-i}^{WR}+\alpha_1}{\sum_{w'} C_{w'r,-i}^{WR}+W\alpha_1} \quad (14)$$

$$P(x_i=1,z=k|\mathbf{w},\mathbf{x}_{-i},\mathbf{z}_{-i},a,\alpha_2,b,c)$$
$$\propto \frac{C_{r2,-i}^{RX}+c}{\sum_{x'} C_{rx',-i}^{XR}+b+c} \cdot (C_{rk,-i}^{RZ}+a) \cdot Multi(f_p|z_i=k) \quad (15)$$
$$\cdot \frac{C_{wk,-i}^{WZ}+\alpha_2}{\sum_{w'} C_{w'k,-i}^{WZ}+W\alpha_2}$$

Since the generation of visual feature is affected by text words, the topic assignment of word should consider the topic assignment of image patches. In Eq. (15), $Multi(.)$ denotes how likely the topic assignment of $w_{p,i}$ matches the topic assignment of image patches, calculated as following:

$$Multi(f_p|z_i=k) = \prod_{j=1}^{N_{f_p}} \frac{C_{rz_i,-i}^{RZ}+a+g(z_{f_{p,j}}=k)}{\sum_{k'} C_{rk',-i}^{RZ}+a+g(z_{f_{p,j}}=k)} \quad (16)$$

where $g(z_{f_{p,j}}=k)$ is an indicator function, equal to 1 if and only if the express inside is evaluated to be true. Equation (15) indicates that the topic assignment for a word is affected by the region preference of the topic and the topic assignments of the patches of the corresponding image.

## 4. LOCATION PREDICTION

To predict the location for a query image, we first find the region that has the greatest joint probability of image content. In reality, a uploaded image may contains only visual content or contains both of visual content and text description. For a query image $q$ with only visual content $f_q$, we selected the target region $\hat{r}$ which has the greatest joint probability as following:

$$\hat{r} = \max_r P(r|\varepsilon)P(f_q|r) = \max_r P(r|\varepsilon) \sum_z P(f_q|z)P(z|r) \quad (17)$$

For a query image $q$ with both text content $w_q$ and visual content $f_q$, the joint probability is revised as following:

$$\hat{r} = \max_r P(r|\varepsilon)P(f_q|r)P(w_q|r) \quad (18)$$

where $P(w_q|r)$ can be estimated with a similar formulation as (6). Then, the location of the query image can be estimated based on the locations of the images located in this region. There are two method to propagate the locations of the images in the target region to the query image. The first method uses the mean location of the selected region as the predicted location directly. This method may be sensitive to outliers. Instead of the mean location, we use the weighted locations of the most similar images in the selected region as the predicted location in the second method. That is, the location of a query image $q$ is estimated as following:

$$l_q = \frac{\sum_{g \in N_n(q)} wei(g) \cdot l_g}{\sum_{g \in N_n(q)} wei(g)} \quad (19)$$

where $N_n(q)$ denotes the $n$ most similar images selected from region $\hat{r}$, and $wei(g)$ is a weight which determines how strongly the result is influenced by image $g$. The similarity between two images is estimated by using JSD (*Jensen-Shannon-Divergence*) to measure the similarity between their two probability distributions over topics which can be obtained from the topic count matrices, i.e., $sim(q,g) = \exp(-D_{js}(P(z|q),P(z|g)))$. The weight is determined by the coherence of image $g_i$ with other images. Since the $n$ images are selected based on their similarity with the query image rather than randomly selected, many of these images should cover the same semantic with the query image and they are more coherent with each other. However, the noisy ones of the top-$n$ images are relatively different from other images, and it is less coherent with other images. Based on this intuition, we estimate $wei(g_i)$ as following:

$$wei(g) = \frac{\sum_{y \in N_n(q)/g} sim(g,y)}{n-1} \quad (20)$$

Equation (20) indicates that, if a image is more similar to other ones of the top-$n$ images, it is more important in determining the location of the query image.

## 5. EXPERIMENTS

In this section, we study the effectiveness of our approach on a real-world dataset.

### 5.1 Dataset

The evaluation dataset is a set of images released by the MediaEval2012 Placing Task [21]. MediaEval2012 contains geo-tagged Flickr (http://www.flickr.com/) images randomly sampled with a method that attempts to maintain coverage of the globe. Since this released dataset includes only the metadata and not the images themselves, we download the raw images using the links in the metadata. Because some images were removed after the dataset was collected, we download one million of images to evaluate the performance. We divided the dataset into 90% for training and 10% for testing. We resize all the images to the resolution of 320*240 pixels and set the maximal patch size $M_1$=300 [20]. For each image, we obtain about 40 to 50 patches. The SIFT feature [14] is used to represent the visual content of each patch. Specifically, we sample a number of keypoints per image, and all the SIFT features of these keypoints are clustered to create a "visual vocabulary". We set the number of clusters to be 1000, and each patch is represented by a 1000-dimensional vector. Textual words are extracted from image's tags, title, and description after filtering the stop words. Some tags contain more than one word. To avoid destroying the semantic information of these tags, we take the whole tag as a single word token.

### 5.2 Baselines and evaluation measure

Here, we are to study the performance of GTMI on image location prediction. Other four approaches are used as baselines in this set of experiments.

The first two approaches are text-based ones, i.e., LGTA [3] and LMSS (language model and similarity search) [16]. We realize LGTA by two steps. First, the region index which maximizes the query image likelihood is selected. Then, the mean location of the region is used as the predicted location. As for LMSS, we use the hybrid method proposed in [16].

The other two approaches are vision-based ones, i.e., IM2GPS [1] and GVR [11]. IM2GPS uses the visual feature distance to find the 130 nearest neighbors and derive the location from these geo-tagged neighbors. To estimate the location of the query image, mean-shift [10] (scale=0.00001) is used to cluster the neighbors based on the location information. Then, the mean location of the cluster which has the highest cardinality is used as the location of the query image. GVR first retrieves a set of top-$K$ visual-neighbors as candidates. Then the candidate whose geographical neighbors that are also in the candidates set and is visually similar to the query image is selected as the target, and the location of the target image is used as the location of the query image.

In practice, image might be uploaded with or without text description. Therefore, we test the performance of location prediction for both types of query image, i.e., query image containing text content and query image without text content, respectively. The two implementations are named GTMI-TV and GTMI-V respectively. For performance evaluation, we calculate the Euclidean distance between the predicted location and real location, using the metric of average distance error (ADE) calculated as following:

$$ADE = \frac{1}{N} \sum_{i=1}^{N} dis(\hat{l}_i, l_i) \quad (21)$$

where $N$ is the total number of query images, and $dis(\hat{l}_i, l_i)$ is the Euclidean distance between the predicted location and the true location. Moreover, the percentages of distances within different ranges are also taken for performance evaluation.

### 5.3 Experimental results

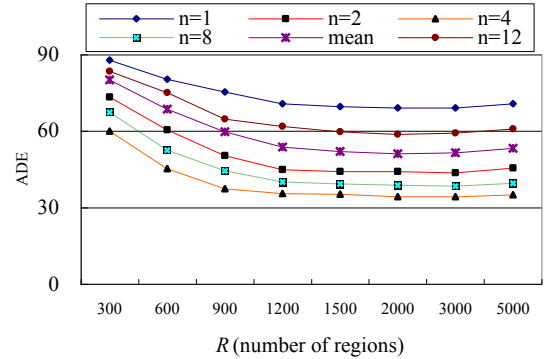

**Figure 2. ADE of the query image containing both text content and visual content with different setting of $n$ and $R$**

There are several parameters in our approach, i.e., $R$-the number of regions used to cluster the training images, and $K$-the number of latent topics in GTMI, $n$-the number of most similar images selected from the target region. First, we fix the topic number to be 100 and evaluate the performance of our approach with different values of $n$ and $R$. Fig. 2 and Fig. 3 show the average distance errors that are obtained for a variety of cluster sizes and number of similar images to consider, where the *mean* line denotes the performance of setting the location of the query image with the mean location of the target region. In Fig. 2, we show the performance of GTMI-TV that predict the location using both of the visual content and text content of the query image, In Fig. 3, we evaluate the performance of another implementation of our approach, i.e., GTMI-V that uses only the visual content of query image to predict its location. From these figures, we find that the performance is not always proportionate to the number of similar images to be selected. The optimal choice of $n$ is 4 for the query image containing both visual content and text content, and 8 for the query image containing only visual content. Setting with the mean location is worse than the method propagating location from similar images to the query image except $n$=1. This is because that the mean location of a region is effect by the outliers. All of the performances are affected by the number of regions that are clustered. If the number is too small, the region will be too coarse, and each region will cover a great number of geo-tagged images many of that may be noisy images. These noisy images may be selected to estimate the location of the query image, and thus the performance decreases. If the number is too large, the chance of selecting a wrong region which has the greatest joint probability of the query image increases, and the time complexity also increases. Therefore, there is trade-off found in figure 2 and figure 3. It can also be concluded that combining different types of content to predict location has a better performance than using only the visual content to predict location, which is because that image text words are related to the geographical location of image greatly. Then, we conduct experiments to analyze how the performance is affected by the parameter of topic number. As LGTA is also based on a topic model, we compare GTMI-TV against LGTA with different numbers of topics and a fixed region number of 1500. As shown in Fig. 4, the average distance error does not change greatly as the number of topics varies. It might because that LGTA predict the location based on the mean vectors of latent regions. Fixing the number of regions is approximate to fixing the range of regions. Thus, a fixed number of regions would bound the prediction performance to some range. Our approach also has a dependence on the result regions. Second, it is clear that GTMI significantly outperforms LGTA. It is because that GTMI learns special words for different regions, which are helpful to discriminate different regions. Moreover, the visual features as well as their relations with textual contents are exploited in GTMI, which is complementary to geographical language model mining. As different region has its own vision patterns and great vision variety, the language model is inadequate to discriminate the geographical characteristic of different regions, which might explain why LGTA performs worse.

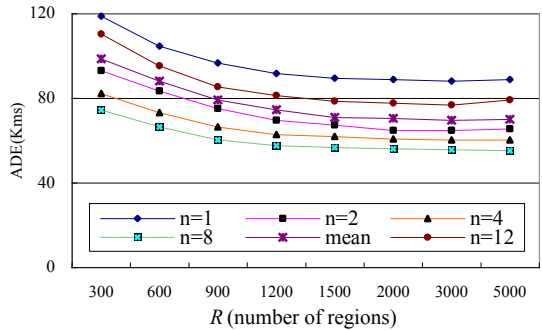

**Figure 3. ADE of the query image containing only visual content with different setting of *n* and *R***

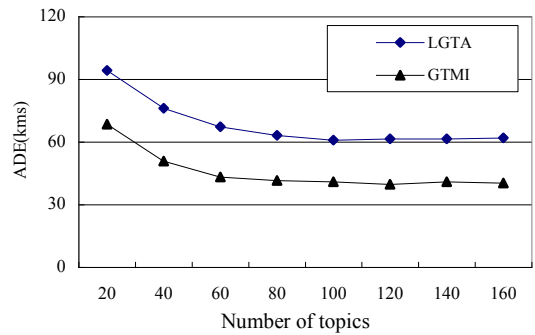

**Figure. 4. Performance comparison with different number of topics**

Previous experiments are used to evaluate how the performance is effected by the parameters setting. In this set of experiments, we compare our approach with other approaches by fixing $n=4$ ($n=8$ for GTMI-V), $K=100$, and $R=1500$. Fig. 5 shows the average distance error of different approaches, with the training data used in these approaches varied from 20% to 100%. It can be concluded that all the performances are affected by the volume of training data. This is might because that when the training dataset is too small, the distributions of images in many locations are very sparse, especially in the locations that are less frequently photographed. Therefore, the prediction of these locations is less effective. It is interesting to find that the text-based approaches perform better than the vision-based approaches. This is because that text content is more effective in conveying the semantic information than the visual content. The result shows that our approach GTMI-TV consistently outperforms other approaches, and GTMI-V perform better than other vision-based approaches. The performances of both IM2GPS and GVR have a high dependency on the selection of visually similar images. However, the visually similar images might be semantically dissimilar images due to the "semantic gap" problem. Thus, visually similar images could be far away from each other. Our approach GTMI-V can discover the geographical discrimination of visual feature based on its correlation with text content and geographical information in the training process. Thus, GTMI-V perform better than these visual similarity based methods, and it performs comparably against other text-based approaches LGTA and LMSS. LGTA and LMSS use the pure text model to discover the geographical language characteristic of each region based on the text content of geo-tagged images, which ignore the large variety of visual content in each geographical location. Therefore their performance is affected. Our approach GTMI-TV integrates different types of contents, using their correlations to identify the latent relation between locations and image's textual content and visual content, which is more effective in location prediction.

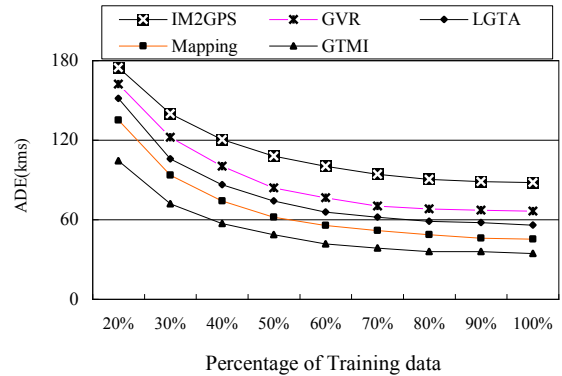

**Figure. 5. Performance comparison with different percentage of training data**

In fig. 6, the percentage of distance errors belonging to different ranges is used as the evaluation measure. It shows the comparison of our approaches against other approaches, with the distance error range varied from 10 Kms to 150 Kms, which indicates that GTMI-TV has the greatest percentages in the former distance ranges compared with other approaches, e.g., 1-10(Kms), 10-30(Kms). Therefore, more number of the query images are predicted by GTMI-TV more accurately, which denotes that GTMI-TV has the lowest ADE. As for GTMI-V, its percentages corresponding to the former distance ranges are greater than other vision-based approaches, and is similar to the text-based approaches. This figure also indicates that the text-based approaches are more effective than the vision-based approaches, since the text content usually is more effective in conveying the semantic information.

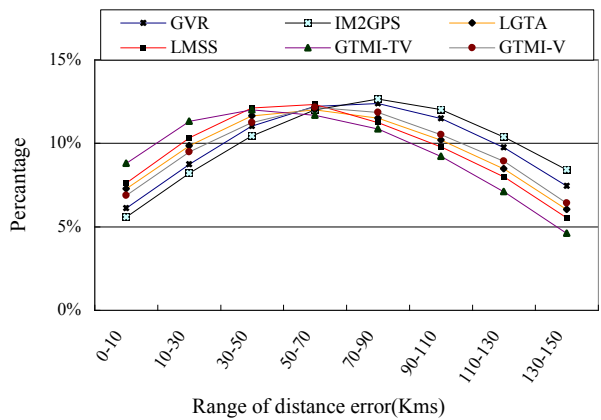

**Figure. 6. Model comparison with different ranges of error**

## 6. CONCLUSION

The emerging trend of geo-tagged social image stimulates a wide variety of novel researches and applications. In this paper, we address the problem of image location prediction by introducing a geographical topic model of social image (GTMI), which simultaneously incorporates multiple types of image contents, i.e., textual description, visual contents, and location information. GTMI introduces the topic structure to combine both text features and visual features, and the latent relation between image content and geo-location is captured by the coherence of topic distributions between image content and regions. A real-life datasets and several baselines are used for comparative studies. Experimental results show that GTMI is effective in location prediction, for new images. Specifically, GTMI is more effective in predicting location for the query image with only visual content than other vision-based approaches, and it is more effective than other text-based approaches in predicting location for the query image with both text content and visual content.

## 7. ACKNOWLEDGMENTS

This work was supported by the National Natural Science Foundation of China (No. 61202239, No. 61170189, and No. 60973105), the Fundamental Research Funds for the Central Universities (No. YWF-14-JSJXY-16), the Opening Project of Beijing Key Laboratory of Internet Culture and Digital Dissemination Research (NO. ICDD201403), and the Fund of the State Key Laboratory of Software Development Environment (No. SKLSDE-2015ZX-11)